\documentclass[aps,pre,twocolumn,superscriptaddress,showpacs,nofootinbib]{revtex4}
\usepackage{graphicx}
\usepackage{amssymb}

\begin{document}
\bibliographystyle{apsrev}

\renewcommand{\thefootnote}{\alph{footnote}}

\title{Nanocapillary adhesion between parallel plates}
\author{Shengfeng Cheng}
\email{chengsf@vt.edu}
\affiliation{Department of Physics, Center for Soft Matter and Biological Physics, and Macromolecules Innovation Institute, Virginia Polytechnic Institute and State University,\\
Blacksburg, Virginia 24061, USA}
\author{Mark O. Robbins}
\email{mr@jhu.edu}
\affiliation{Department of Physics and Astronomy, Johns Hopkins University,\\
3400 N. Charles Street, Baltimore, Maryland 21218, USA}

\date{\today}

\begin{abstract}
Molecular dynamics simulations are used to study capillary adhesion from 
a nanometer scale liquid bridge
between two parallel flat solid surfaces.
The capillary force, $F_{cap}$, and the
meniscus shape of the bridge are computed as 
the separation between the solid surfaces, $h$, is varied.
Macroscopic theory predicts the meniscus shape and 
the contribution of liquid/vapor interfacial tension to $F_{cap}$
quite accurately for $h$ as small as 2 or 3 molecular diameters (1-2 nm).
However the total capillary force differs in sign and magnitude from
macroscopic theory for $h \lesssim 5$ nm (8-10 diameters)
because of molecular layering that is not included in macroscopic theory.
For these small separations, the pressure tensor in the
fluid becomes anisotropic.
The components in the plane of the surface vary smoothly and are consistent
with theory based on the macroscopic surface tension.
Capillary adhesion is affected by only the perpendicular component,
which has strong oscillations as the molecular layering changes.
\end{abstract}


\maketitle

\section{Introduction}
Capillary adhesion from liquid bridges between solids allows us to build
sandcastles, enables insects to stick on a ceiling, and
causes granules to agglomerate.\cite{hornbaker97,weiss00,rabinovich02,herminghaus05,scheel08}
Condensation induced capillary adhesion is common whenever hydrophilic
surfaces are in a humid environment\cite{riedo02,charlaix10} and is a
major cause of failure in microelectromechanical systems.\cite{weiss00,maboudian02}
Like other interfacial forces, capillary adhesion grows
in importance as dimensions shrink to molecular scales.
However capillary forces are typically modeled using macroscopic theory
that must fail in the same limit.\cite{rowlinson89}
In this paper we use molecular dynamics (MD) simulations to explore molecular scale changes in capillary adhesion in a simple geometry, a liquid bridge between parallel plates.

\begin{figure}[htb]
\centering
\includegraphics[width=3in]{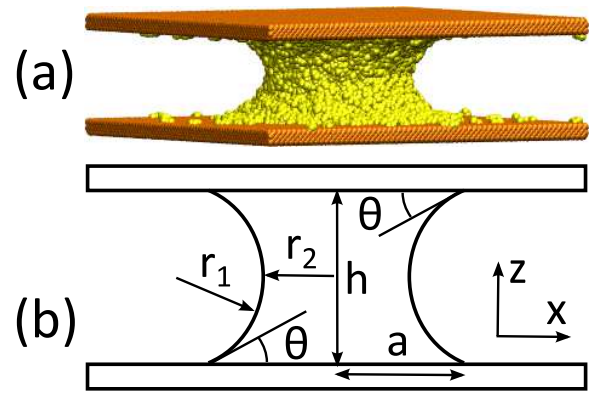}
\caption{(a) A liquid bridge (yellow) connects two parallel atomically flat plates (orange). 
(b) Geometry is specified by the contact angle $\theta$ and the radius $a$ 
of the contact line at which the bridge intersects with the plates.
The radius of curvature is negative (positive) when the liquid/vapor interface 
bends outward (inward) in the plane of curvature,
as illustrated for $r_1$ ($r_2$).}
\label{AtomicConfig}
\end{figure}

Macroscopic theory describes capillary forces in terms of two contributions.\cite{rowlinson89,lambert08}
One is due to the surface tension from the interface of the drop 
(Fig. \ref{AtomicConfig}).
The second comes from the area times the Laplace pressure in the drop
due to the curvature of the interface.
As dimensions shrink to nanometer scales, both terms may become inaccurate due to the finite width of the interface, changes in surface tension, changes in meniscus geometry or other new phenomena.

Some previous research attempts to identify the limits of macroscopic equations of capillarity
at small length scales.\cite{fisher79,fisher81a,fisher81b,kohonen00,thompson93b,bresme98,takahashi13}
For example, it was shown that the Kelvin equation of capillary condensation, 
which relates the pressure difference across
a liquid/vapor interface to the vapor pressure, is obeyed by cyclohexane and water menisci
with a mean radius of curvature as small as $4$ nm.\cite{fisher79,fisher81a,fisher81b,kohonen00}
The Young-Laplace equation, which relates the the pressure difference 
between the liquid and vapor phase to the interfacial tension and the mean curvature of the interface,
is found to be valid down to a similar scale.\cite{thompson93b,bresme98,takahashi13}
However, it is not obvious if macroscopic theory can describe capillary
forces due to liquid menisci spanning small gaps between solid surfaces,
such as nanoscopic slits in porous media and granular materials
or between a probe and a solid surface.

In a previous paper we used molecular dynamics (MD) simulations to study
the capillary adhesion induced by
a liquid bridge between a spherical tip and a flat substrate.\cite{cheng14pre}
At small separations the capillary force ($F_{cap}$) became progressively
less attractive than predicted by macroscopic theory and oscillated rapidly
with surface separation $h$.
Analysis of the local pressure showed that these oscillations were
due to layering of molecules.
The layered structure led to an anisotropic pressure-stress tensor.
The normal pressure component determines the term in $F_{cap}$ from the integral
of pressure over area.
It developed rapid fluctuations superimposed on a gradual decrease in adhesion that explained the deviations from macroscopic theory.
The in-plane component remained consistent with macroscopic theory as did the detailed shape of the meniscus.
However it was difficult to probe the interfacial contribution in very
narrow gaps because this work used a sphere on flat geometry.
The separation between solids approached molecular scales in the center of the drop, but increased significantly by the outer meniscus.
A better test of the interfacial contribution to $F_{cap}$ requires a different geometry.

In this manuscript we consider the case of two parallel plates that sandwich
a liquid bridge in a uniform gap. This geometry allows us to 
further check the limits of the macroscopic capillary theory based on the Young-Laplace equation.
Figures \ref{AtomicConfig}(a) and (b) show a snapshot and a sketch of the system.
This geometry approximates experiments where a small meniscus is
trapped between two surfaces with very large radii of curvature.
Examples include the gap between a blunt tip and a flat surface, 
between two large granular particles, or between two cross-aligned mica surfaces in a Surface Force Apparatus (SFA).
Such geometries are of growing technological relevance 
because of the important role played by fluid bridges and joints in micro- and nanosystems
that may arise from condensation, excretion, or trapping of liquid.\cite{mastrangeli15}

As in our previous study we find that deviations from macroscopic theory are predominantly due to changes in the normal pressure caused by molecular layering.
The interfacial contribution remains consistent with macroscopic theory down to gaps that are only 4 atoms across.
At still smaller scales the capillary force becomes more attractive because the interface has too few atoms to approximate the predicted curvature of the meniscus.
However the in-plane component of the pressure tensor continues to follow macroscopic theory to the smallest scales studied.

In the next section we provide a brief description of the simulation technique,
followed by the equations of macroscopic theory.
The following section presents our results for different contact angles and
gaps.
We end with a summary and conclusions.

\section{Methods}

\subsection{Computational model}

Since our goal is to address generic behavior, we use simple potentials that have
been shown to capture many aspects of the behavior of short chain molecules and polymers.\cite{kremer90,kroger93,larson00,rottler03c,kroger04,cheng10pre,cheng14pre}
Fluids are modeled as linear chains of 4 spherical beads.
All beads interact with a truncated Lennard-Jones (LJ) potential
\begin{equation}
\label{LJPotential}
V_{\rm LJ}(r) = 4\epsilon\left[ \left( \sigma/r \right) ^{12}-\left( \sigma/r \right)^{6}
 -\left( \sigma/r_c \right)^{12}+\left( \sigma/r_c \right)^{6} \right]~.
\end{equation}
where $\epsilon$ is the interaction energy, $\sigma$ the molecular diameter, and $r_c$ the cutoff length.
For beads not directly bonded,  $r_c=2.2\sigma$.
For two neighboring beads in a chain molecule, in addition to a purely repulsive LJ interaction (i.e., $r_c=2^{1/6}\sigma$), a finitely extensible nonlinear elastic (FENE) potential
\begin{equation}
\label{FENEPotential}
V_{\rm FENE}(r) = -\frac{1}{2} K R_0^2 {\rm ln}\left[ 1-\left( r/R_0 \right) ^2 \right]~,
\end{equation}
with the canonical values $R_0=1.5\sigma$ and $K=30\epsilon/\sigma^2$ is used to describe the bonded interaction.

The LJ interaction energy $\epsilon$, bead diameter $\sigma$, and mass $m$
are used to define all units.
To map the results to real units we use the facts that a typical hydrocarbon has molecular
diameter $\sigma \sim 0.5~{\rm nm}$ and surface tension $\gamma=25~{\rm mN/m}$.
Since the liquid/vapor interfacial tension computed in simulations
is $\gamma =0.88\epsilon/\sigma^2$, we find $\epsilon \sim 7\times 10^{-21}~{\rm J}$.
Beads typically correspond to a small cluster of atoms\cite{kremer90} with mass $m \sim 10^{-25}$ kg.
Then the characteristic LJ time, $\tau= \sqrt{m\sigma^2/\epsilon}$,
is $\sim 2$ ps.
The unit of force is $\epsilon/\sigma \sim 14$ pN and the unit of pressure
$\epsilon/\sigma^3 \sim 56$ MPa.

We performed MD simulations using the LAMMPS package.\cite{LAMMPS}
The equations of motion are integrated with a time step of $0.005\tau$.
Constant temperature is maintained using a Langevin thermostat with time constant $1.0\tau$.
For the results presented below $T=0.75\epsilon/k_{\rm B}$, where $k_{\rm B}$ is the Boltzmann constant.
This temperature is intermediate between the temperature $1.0\epsilon/k_{\rm B}$ typically used for melt
simulations\cite{kremer90} and the glass transition temperature $\sim 0.4 \epsilon/k_{\rm B}$.\cite{buchholz02,rottler03c} 
Given our estimate of $\epsilon$, the temperature $0.75\epsilon/k_{\rm B}$ maps to a reasonable value of $\sim 360$K.
Lowering the temperature to $0.5 \epsilon/k_{\rm B}$ did not change the trends reported
below but did increase the degree of layering and the magnitude of force oscillations.

For simulations reported here, the plates are treated as rigid bodies.
We found that including elasticity had a negligible effect\cite{cheng14pre} because any deformation of the
solids is much less than $\sigma$ for elastic moduli typical of molecular solids.
\footnote{Elastomers have smaller moduli but also a much more complicated local structure
than that considered here.\cite{style12,style13}}
Each plate is modeled as an fcc crystal with a (001) surface and number density $1.0 \sigma^{-3}$.
The fluid beads interact with the solid atoms via a LJ potential with 
modified interaction energy $\epsilon_{fs}$, length scale $\sigma_{fs}=1.2\sigma$,
and cutoff $r_{cfs}=2.16\sigma$.
The interaction strength $\epsilon_{fs}$ was varied to control the contact angle $\theta$, which
was calculated by placing a drop on the plate
and fitting the equilibrated drop shape to a spherical cap.\cite{cheng14pre}
We found $\theta=75^\circ$ for $\epsilon_{fs}=0.8\epsilon$,
while $\theta$ is reduced to $12^\circ$ 
when $\epsilon_{fs}$ is increased to $1.08\epsilon$.
We confirmed that any effects due to line tension are negligible.\cite{amirfazli04,cheng14pre}

Care must be taken in defining the plate separation that corresponds to macroscopic
theory.
We define $h_a$ as the distance between the closest atoms on the surfaces of the
opposing solid plates.
This overestimates the volume available to the fluid because of the steric repulsion between fluid and solid atoms.
To determine the effective width $h_{ex}$ of the excluded volume near each plate
we performed simulations of a fluid that contains $N$ molecules and fills the space
between two parallel solid plates of area $A$ at zero pressure.
The value of $h_a$ should give an accessible thickness $h \equiv h_a-2h_{ex}$
equal to that expected for the bulk density $\rho_b$ at zero pressure:
$h= h_a-2h_{ex}=N /(A \rho_b)$.
We found $h_{ex}=0.775\sigma$ for $\epsilon_{fs}=0.8\epsilon$ and $h_{ex}$ decreases to 0.710$\sigma$ for $\epsilon_{fs}=1.08\epsilon$.
Below we use $h$ as the separation but the value of $h_a$ differs only by a constant offset for a given $\epsilon_{fs}$.

To form a liquid bridge, a drop was initially deposited on the surface of the bottom plate and 
allowed to relax to its equilibrium configuration. 
Then the top plate was brought down to contact the drop, creating a liquid bridge between the two plates. For the small contact angle shown in Fig. \ref{AtomicConfig}(a), a few of the molecules escape along the surface when the meniscus forms.
The number remains small ($< 0.3\%$) and the vapor pressure is so low that no molecules
evaporate.
This justifies the use of a constant volume ($V_l$) ensemble in comparing
macroscopic theory to simulation results.
We have studied bridges with various volumes and all show
similar trends. In this report we focus on a bridge with
$9,316$ molecules (37,264 beads). 
This corresponds to $V_l = 4.123\times 10^4 \sigma^3$
at the equilibrium bulk density at zero pressure, $\rho_b = 0.904 \sigma^{-3}$.
There are small changes in density and volume with Laplace pressure that can be
included in macroscopic theory.
For the largest pressures found here ($\sim 0.5 \epsilon/\sigma^3$)
the changes in density are at most 1\%, leading to changes in the predicted
$F_{cap}$ of the same order.

The separation between plates was varied in small steps.
After each step, the liquid bridge was allowed to relax for at least $2000\tau$ before
the local and global forces were calculated.
The equilibration time of the liquid bridge was less than $1000\tau$ for $h \gtrsim 2\sigma\sim 1~{\rm nm}$, and we found negligible hysteresis in the forces.\cite{cheng14pre}
Hysteresis between increasing and decreasing separations was only found when
$h < 2\sigma$ and the film was in a glassy state.
For this reason we only present results for $h \gtrsim 2\sigma$ where the liquid bridge has reached equilibrium.

\subsection{Macroscopic theory of capillary forces}

In the macroscopic theory of capillary phenomena,
the shape of the liquid bridge is determined by the Young-Laplace equation.\cite{rowlinson89}
A sketch of the geometry is shown in Fig. \ref{AtomicConfig}(b).
The interface must intersect the plates at the equilibrium contact angle $\theta$
given by Young's equation
and obey the Young-Laplace equation for the local pressure change $\Delta p$ across the curved interface
\begin{equation}
\Delta p = \gamma (1/r_1 + 1/r_2) \equiv 2 \gamma \overline{\kappa}~,
\label{Laplace}
\end{equation}
where $r_i$ are the principal radii of curvature and $\overline{\kappa}$ is the mean curvature.
The radius is positive (negative) when the center of the circle that touches the interface is inside (outside) the bridge.
In Fig. \ref{AtomicConfig}(b) $r_1$ is negative.
The in-plane radius $r_2$ is always positive.

Since gravity is negligible for nano-sized liquid bridges, $\Delta p$ and thus $\overline{\kappa}$
must be constant for a given bridge.
For nonvolatile liquids, $\Delta p$ is fixed
by the volume $V_l$ of the bridge.
For volatile liquids, $\Delta p$ is determined
by the relative humidity of the vapor via the Kelvin equation.\cite{rowlinson89}
The results reported here are for a nonvolatile liquid because 
the liquid composed of chain molecules evaporates extremely slowly.\cite{cheng11jcp}
Results for volatile liquids are the same for a given $V_l$ and $h$, but have different variations with $h$ since $V_l$ changes.

The capillary force has two terms in the macroscopic theory,\cite{rowlinson89,lambert08}
\begin{equation}
F_{cap}= -2\pi a \gamma \sin\theta + \pi a^2 \Delta p~,
\label{force}
\end{equation}
where $a$ is the radius of the contact circle at which the bridge intersects with the solid surface.
Our sign convention is that negative (positive) values correspond to attractions (repulsions).
The first term is the vertical projection of the surface tension force
and is always attractive, since moving the solid surfaces closer
reduces the area of liquid/vapor interface and thus the surface free energy.
This term is referred to below as $F_s$.
The second term is the integral of the Laplace pressure
over the circle where the fluid contacts the solid.
This term can be either attractive or repulsive, depending on the sign
of $\Delta p$, and is designated as $F_p$.

Equation (\ref{Laplace}) can be solved exactly using elliptic integrals.\cite{orr75}
The resulting continuum predictions for the shape of the bridge,
$F_s$, $F_p$, and $F_{cap}$ are then
compared to MD results for the same $\theta$ and $V_l$.

\section{Results and discussion}

\begin{figure}[htb]
\centering
\includegraphics[width=3in]{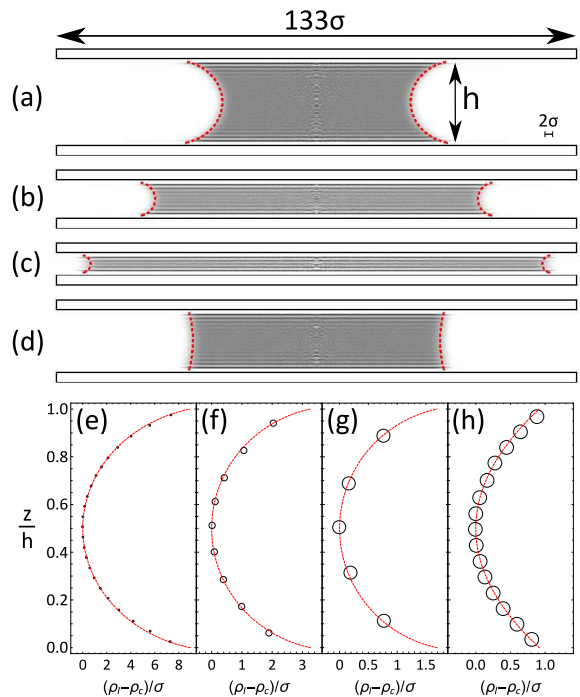}
\caption{
Comparison of the interfacial shape from simulations and macroscopic theory (dashed red lines)
for $\theta=12^\circ$
at (a,e) $h=19.9\sigma$, (b,f) $h=7.9 \sigma$, 
(c,g) $h=4.1\sigma$
and for 
$\theta=75^\circ$ at (d,h) $h=13.7 \sigma$.
In (a)-(d) the angular averaged density $n(\rho,z)$ is represented by a gray scale plot.
A horizontal scale bar indicates a width of $2\sigma$.
In (e)-(h), circles indicate the interface position $\rho_I$
relative to the radius at the center of the meniscus $\rho_c$.
}
\label{InterfaceShape}
\end{figure}

Within statistical fluctuations, droplets have the circular shape expected from symmetry and predicted by macroscopic theory.
Figures~\ref{InterfaceShape}(a)-(d) show the angle-averaged density profile $n(\rho,z)$ as a function of height and radial distance $\rho$ from the center of the bridge.
In all cases, there are clear oscillations in $n$ with height that
reflect layering in the film.\cite{nordholm80,toxvaerd81,magda85,bitsanis90a,gao97c,thompson90a,cheng14pre}
The LJ potential favors a spacing of order $\sigma_{fs}$ between solid and fluid atoms.
A sharp solid wall induces a layer of fluid at the equilibrium spacing.
This layer then induces a second layer at the equilibrium spacing between fluid beads.
Past studies of homogeneous thin films without menisci show that layers decay exponentially with height.\cite{nordholm80,toxvaerd81,magda85,bitsanis90a,gao97c,thompson90a,cheng14pre}
The same behavior is evident in Figs. \ref{InterfaceShape}(a)-(d) at small $\rho$.
There are always large oscillations in density near the wall, but this layering only spans the entire film
for panels (b) and (c) where $h < 8\sigma$.

The interfaces of the drops in Figs.~\ref{InterfaceShape}(a)-(d) are broadened by molecular discreteness
and thermal fluctuations.
At each $z$, the width of the interfacial region where $n$ changes
is comparable to molecular dimensions and a horizontal $2\sigma$ scale bar is included
for reference.
Also shown in Figs.~\ref{InterfaceShape}(a)-(d) is the macroscopic prediction for the interface shape
(red dashed lines).
Even at the smallest wall spacings the macroscopic predictions follow the interfacial region and lie within the
range where $n$ is decaying rapidly with $\rho$.

To make a more precise determination of the interfacial shape and width
we analyzed the decay in $n(\rho,z)$ with increasing $\rho$.
Results were averaged over a height range comparable to the period of density oscillations, $\Delta z \sim 0.8\sigma$,
because $n(\rho,z)$ drops to nearly zero at some heights near the wall.
At small $\rho$, the density has a nearly constant value of $n_0(z)$.
As $\rho$ increases through the interfacial region, $n(\rho,z)$ drops to zero.
There is no unique definition for the interface position within the interfacial region,\cite{rowlinson89}
but a reasonable choice is the radius where $n(\rho,z) = 0.5 n_0(z)$.
Values were obtained by fitting to a common analytic form
for liquid-vapor interfaces,\cite{jasnow96}
$n(\rho,z)=\frac{1}{2}n_0(z) (1- {\rm tanh}\frac{\rho - \rho_I}{\sqrt{2}\xi})$,
where $\rho_I$ is the interface location and $\xi$ the interface half-width at a given $z$.
Values of the halfwidth were in the range $\xi =1.2 \pm 0.4$ and depended
on both $h$ and $z$.
As shown in previous work, $\xi$ reflects both an intrinsic interface halfwidth and an additional broadening from thermal fluctuations.\cite{denniston04,lacasse98}

Circles in Figs.~\ref{InterfaceShape}(e)-(h) show the interface
position $\rho_I$ at heights corresponding to the centers of layers
near the wall or in central regions where $n$ is nearly independent of height.
Red lines show the corresponding macroscopic prediction.
In both cases $\rho_I$ is referenced to the radius at the center of the film $\rho_c$.
In all cases studied, the shapes of the interface
from simulations and macroscopic theory are consistent within statistical
fluctuations, which are indicated by the radii of circles ($\sim 0.1\sigma$).
Thus the curvatures that enter the Laplace equation for the pressure are
accurately predicted by macroscopic theory.
One caveat is that the number of layers where the interface
is defined decreases as $h$ is reduced.
Effects from this discreteness at $h<4\sigma$ are noted below.

\begin{figure*}[htb]
\centering
\includegraphics[width=6in]{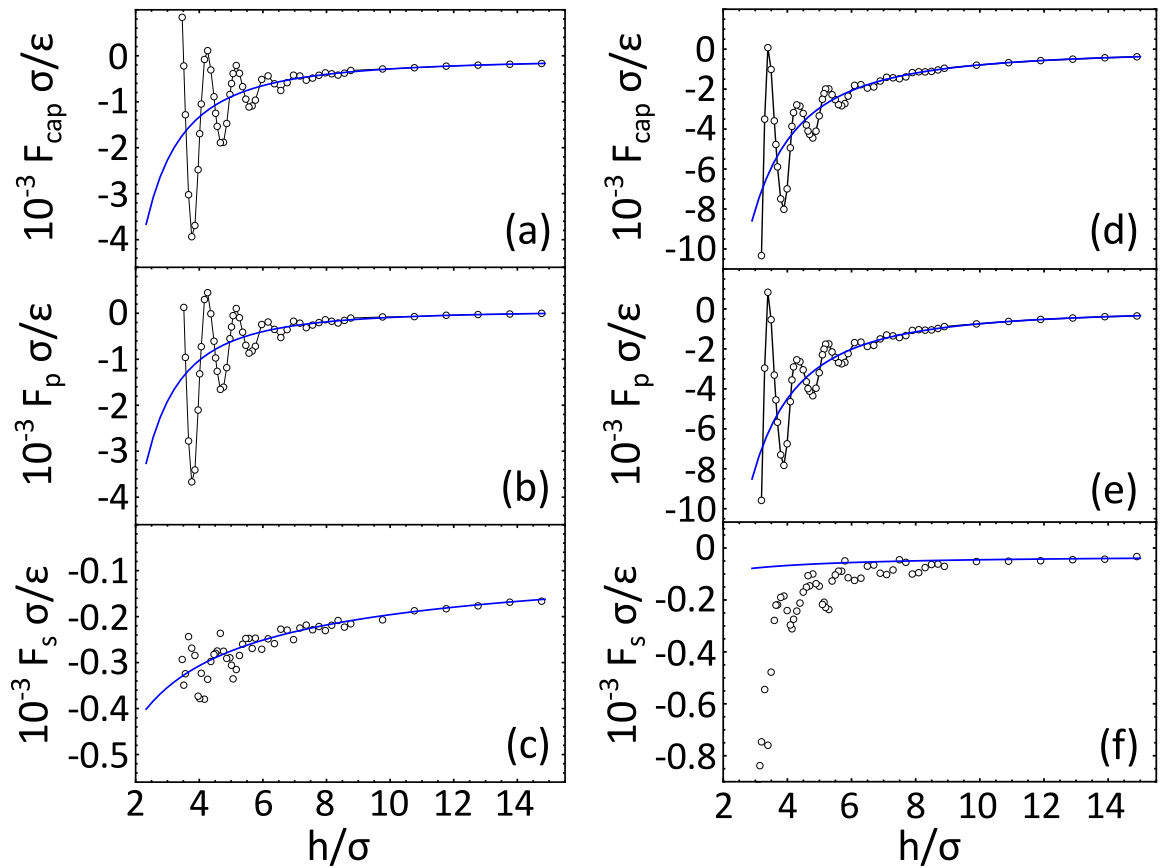}
\caption{Capillary forces vs. $h$: (a) and (d) total force;
(b) and (e) contribution from the Laplace pressure;
(c) and (f) contribution from the liquid/vapor surface tension.
(a)-(c) are for $\theta=75^\circ$ and (d)-(f) for $\theta=12^\circ$, respectively.
Blue curves represent the corresponding continuum predictions.
Lines connecting data points are guides for the eye.}
\label{ForTotPlot}
\end{figure*}

Figures \ref{ForTotPlot}(a) and (d) show $F_{cap}$ vs. $h$
for $\theta=75^\circ$ and $\theta=12^\circ$, respectively.
In both cases good agreement is found between the simulations and 
continuum predictions for $h \gtrsim 8\sigma \sim 4$ nm.
As $h$ decreases below $8\sigma$, pronounced oscillations in $F_{cap}$ become apparent.
These oscillations reflect molecular layering in the gap between plates.\cite{nordholm80,toxvaerd81,magda85,bitsanis90a,gao97c,denniston06,cheng14pre}
As noted above, layering is evident in Fig.~\ref{InterfaceShape} and clearly
spans the film for $h < 8\sigma$.

Similar oscillating forces were first observed in SFA
experiments\cite{horn81,israelachvili84}
and later found in atomic force microscopy measurements.\cite{ jarvis00,kocevar01}
However in these experiments the entire space was filled with fluid and the oscillations
represent a variation in the free energy of the film as a function of 
thickness rather than the capillary force.
The derivative of the free energy per unit area with respect to thickness is called 
the disjoining pressure and its integral over the solid surfaces gives the net force
between surfaces in analogy to the pressure contribution in Eq.~\ref{force}.

Forces induced by molecular layering are attractive when $h$ is close to but larger
than an integral multiple of the equilibrium layer spacing because
the free energy is lowered by decreasing the spacing.
The force becomes repulsive when $h$ is reduced below the optimal spacing
and becomes attractive again when a layer of molecules is pushed out and
$h$ approaches the next integral number of layer spacings (see the inset of Fig.~\ref{StressPlot} below).
This cycling process leads to the oscillating behavior of $F_{cap}$
as shown in Figs.~\ref{ForTotPlot}(a) and (d).

\begin{figure}[htb]
\centering
\includegraphics[width=2.5in]{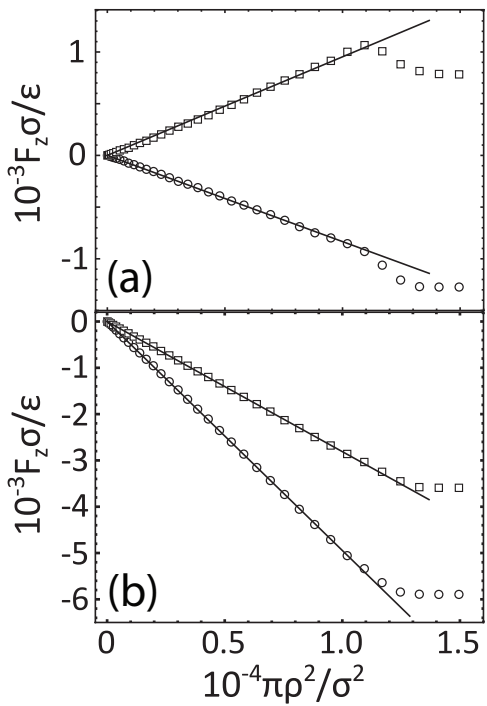}
\caption{The integral of the vertical force $F_z$ within a circle of radius $\rho$ 
vs. $\pi\rho^2$ at $h_a=5.1\sigma~(\bigcirc)$ and $5.0\sigma~(\Box)$ for 
(a) $\theta=75^\circ$ and (b) $\theta=12^\circ$.
Each linear fit gives a slope equal to the normal pressure $P_n$.}
\label{ForAreaPlot}
\end{figure}

In the macroscopic theory, $F_{cap}$ is the sum of the surface tension term $F_s$ and 
the pressure term $F_p$.
To separate the two terms in simulations,
we examine the vertical component of the local force between the
fluid and solid atoms as a function of the lateral distance $\rho$
from the center of the bridge.
The integral $F_z$ within a circle of radius $\rho$ is plotted
against the circle's area $\pi \rho^2$ in
Fig.~\ref{ForAreaPlot} to reduce noise from thermal fluctuations.
The linear relationship between the total force and area at small $\rho$
indicates a uniform normal pressure at the solid-liquid interface that
determines $F_p$.
Near the contact line the integral drops below this linear slope,
reflecting the attractive contribution $F_s$
from the surface tension of the liquid/vapor interface.

Linear fits like those in Fig.~\ref{ForAreaPlot} 
allow us to extract a slope corresponding to the normal pressure 
$P_n=\frac{1}{\pi}\frac{\partial F_z}{\partial \rho^2}$
inside the liquid bridge.
Note that $P_n$ changes vary rapidly with $h$.
In both examples shown, decreasing $h$ by $0.1\sigma$ changes $P_n$ by more than $0.2 \epsilon/\sigma^3$.
This changes the sign of the force for $\theta=75^\circ$.
The background attraction is stronger for $\theta=12^\circ$, but the
the magnitude of $P_n$ decreases by about a factor of 2.
The oscillations in $P_n$ increase as $h$ decreases and
make the net force repulsive for $\theta=12^\circ$ at $h\simeq 3.4\sigma$ (Fig~\ref{ForTotPlot}(d)).

The pressure contribution to $F_{cap}$ is obtained 
as $F_p= \pi a^2 P_n$ where $a$ is the radius of the contact circle.
We calculated $a$ directly from the area where wall and fluid atoms
interact and also from macroscopic theory.
As expected from Fig. \ref{InterfaceShape}, the two values agreed within
less than 1\% in all cases and we use the theoretical value
in all the results presented below.
The surface tension term is then computed as the remaining force $F_s=F_{cap}-F_p$.

Figures \ref{ForTotPlot}(b) and (c) show $F_p$ and $F_s$ as a function of $h$
for $\theta=75^\circ$.
The results for $\theta=12^\circ$ are included in Figs.~\ref{ForTotPlot}(e) and (f).
For both $\theta$, the pressure term ($F_p$) agrees with the continuum curve for $h>8\sigma\sim 4$ nm.
At smaller $h$, $F_p$ oscillates about the continuum solution
and the magnitude of deviations grows as $h$ decreases.
These oscillations are very similar to those found in the total force.
They account for almost the entire deviation from macroscopic theory
because $F_s=F_{cap}-F_p$ is very close to the continuum prediction
down to $h< 4\sigma$.
For $\theta = 75^\circ$ the discrepancies in $F_s$ at small $h$ are within numerical uncertainties because $F_s$ is calculated as the difference between two much larger oscillating quantities and depends on the exact contact radius used.
For $\theta=12^\circ$ there is a systematic increase in attraction at $h<4\sigma$.

The success of continuum expressions for $F_s$ is consistent
with the agreement between the predicted and simulated shapes of the liquid
bridge (i.e. Fig. ~\ref{InterfaceShape}) and our previous results
for sphere on flat geometries.\cite{cheng14pre}
Clearly the magnitude of the interfacial tension remains equal to the bulk value even in menisci that are only two or three molecules across.
The increase in the magnitude of $F_s$ for $\theta=12^\circ$ at $h < 4\sigma$ can be understood as a geometric effect.
Macroscopic theory predicts a radius of curvature $|r_1| \approx  h/(2 \cos \theta) $.
For $\theta=75^\circ$ this is much larger than $h$.
For $\theta=12^\circ$, $|r_1|\approx h/2$ and
as noted in discussing
Fig. \ref{InterfaceShape}
it is not possible for the interface position at
a few discrete layers to closely approximate the predicted circle.
Indeed, for a two layer system the interface is always nearly vertical,
leading to
a capillary force proportional to $\gamma$ instead of $\gamma \sin \theta$.
This only changes the predicted $F_s$ by about 4\% for $\theta=75^\circ$, but
increases the magnitude of $F_s$ by a factor of almost 5 for $\theta=12^\circ$.
This provides a quantitative explanation of the changes in Fig. \ref{ForTotPlot}(f).

It may be surprising that the fluid pressure from macroscopic theory gives the right interfacial shape from Eq. \ref{Laplace} but is inconsistent with the pressure that determines $F_p$.
The resolution is that the pressure tensor becomes anisotropic at small $h$.
The curvature of the interface is predominantly determined by
the in-plane component $P_{\rho \rho}$, which remains consistent with macroscopic
theory and isotropic within the plane of the plate.
The normal pressure $P_{zz}$ determines $F_p$.
For large $h$ the pressure is hydrostatic and $P_{zz}=P_{\rho\rho}$.
This symmetry is broken as $h$ decreases.
Layering leads to strong oscillations in $P_{zz}$ with $h$, but does not affect $P_{\rho\rho}$.

\begin{figure}[htb]
\centering
\includegraphics[width=3in]{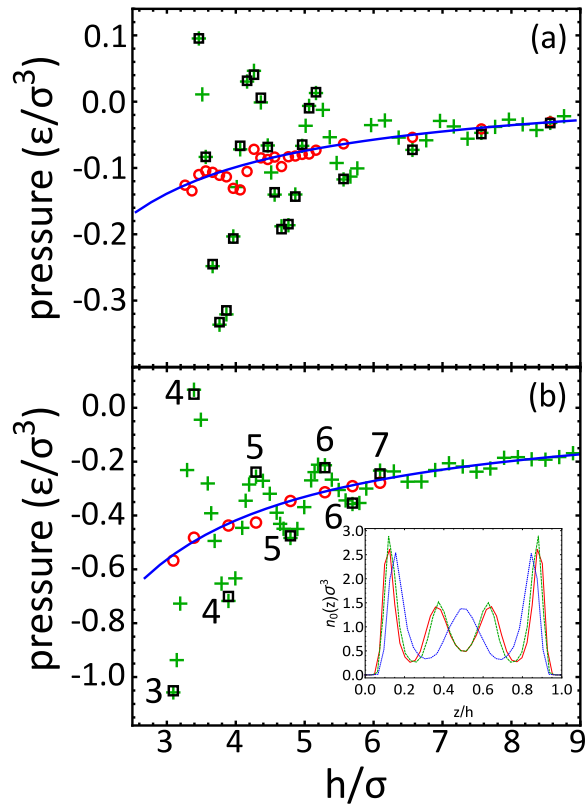}
\caption[Pressure anisotropy in a liquid bridge]
{Variation of the pressure tensor with plate separation
for (a) $\theta=75^\circ$ and (b) $\theta=12^\circ$.
The directly calculated in-plane component
$P_{\rho\rho}$ (red $\bigcirc$)
and out-of-plane component $P_{zz}$ (black $\square$) of the pressure tensor
are compared to
$P_n$ (green $+$) from the linear fits in Fig.~\ref{ForAreaPlot} and
the continuum prediction of the Laplace pressure $\Delta p$ (blue solid line).
Integers in (b) indicate the number of layers at local minima and
maxima in $P_{zz}$. The inset shows density profiles at the
extrema associated with 4 and 3 layers
at $h=3.9\sigma$ (red solid line), $3.4\sigma$ (green dashed line), and $3.1\sigma$ (blue dotted line).
}
\label{StressPlot}
\end{figure}

Figure~\ref{StressPlot} shows direct evaluations of the local pressure tensor
inside the liquid bridge using an algorithm proposed by Todd et al. \cite{todd95}.
Equivalent results were obtained with a different algorithm described
by Denniston and Robbins.\cite{denniston04,denniston01}
Note that the radial, in-plane component $P_{\rho\rho}$ follows
the continuum prediction for $\Delta p$ even in films as thin
as $3\sigma\sim 1.5$ nm.
In contrast, the normal pressure $P_{zz}$ 
oscillates around the continuum prediction for $\Delta p$ 
and the deviations increase as $h$ decreases.
These directly calculated values of $P_{zz}$ are completely consistent
with the normal stress $P_n$ extracted from Fig.~\ref{ForAreaPlot}.
\footnote{Note that while $P_{zz}$ oscillates with $h$, it is independent of height for a given $h$. This is required for any system in equilibrium.}

The pressure anisotropy reflects the influence of surfaces on the liquid under confinement.
The layered structure induced by the surface leads to oscillations with $h$ in the free energy and disjoining pressure that are not present in macroscopic theory.
These lead to the oscillations in normal pressure with $h$ in Fig.~\ref{StressPlot}.
For both values of $\theta$ the oscillations become large enough to change the magnitude of the force by a factor of 2 or change the force from attractive
to repulsive.

The changes in layering are illustrated in Fig.~\ref{StressPlot}(b).
Minima and maxima in the pressure are labelled with the corresponging number of layers.
At each minimum, the spacing is larger than the optimal spacing for the
corresponding number of layers and there is an extra attraction pulling
the surfaces together to the optimum spacing.
As $h$ decreases to the optimal spacing, this attractive term vanishes.
Compressing the layer past the optimal spacing gives a repulsive force
that grows until a layer is squeezed out.
The inset shows density profiles corresponding to the minimum and maximum
for 4 layers and the subsequent minimum after a layer has been squeezed out to leave 3 layers.

\begin{figure}[htb]
\centering
\includegraphics[width=2.75in]{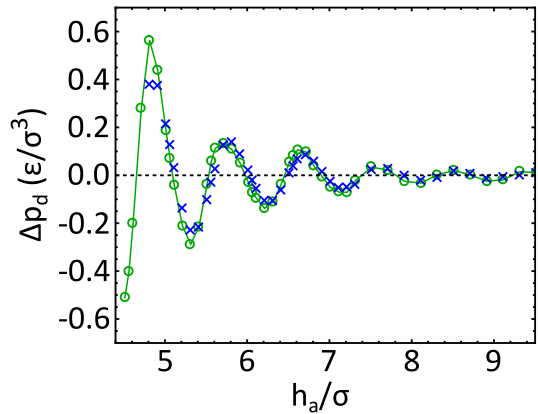}
\caption[Deviation from macroscopic theory]
{The disjoining pressure ($\Delta p_d = P_n-P_{\rho\rho}$) as a function
of the separation between wall atoms $(h_a)$ for $\theta=75^\circ$ (blue $\times$) and $\theta=12^\circ$ (green $\bigcirc$). The lines are guides for the eye. }
\label{DisjoiningPressure}
\end{figure}

The disjoining pressure $\Delta p_d$ is present even when there is no meniscus and should only depend on the local surface separation, surface interactions, density and curvature.
Since the disjoining pressure only produces a contribution to the normal
pressure, we define $\Delta p_d = P_n - \Delta p$ where $\Delta p$ is the
continuum prediction for the Laplace pressure and is consistent with $P_{\rho \rho}$.
As shown in Fig.~\ref{DisjoiningPressure}, $\Delta p_d$ oscillates around
zero and is nearly the same for both contact angles studied here.
Although different contact angles reflect different
interactions with the walls, the interactions are short range.
Most of the variation in free energy comes from entropic packing
effects that are determined by the wall spacing ($h_a$) but independent of $\theta$.
This is the reason that we plot $\Delta p_d$ against $h_a$ in Fig.~\ref{DisjoiningPressure}.
The difference in direct interactions only appear to be important at the smallest $h_a$ where there are roughly 3 molecular layers.
In our previous study of the sphere on flat geometry, there was a repulsive shift in $F_{cap}$ as well as oscillations.
This repulsive shift is not present for flat surfaces and we conclude it
reflects an extra free energy cost of changing the layered structure
as the curvature of the solid sphere causes the gap width to change.

The agreement between $P_{\rho\rho}$ and macroscopic theory extends to $h=3\sigma$.
This is surprising given our finding that $F_s$ becomes more negative due to
the inability of the interface to approximate a sphere on this scale.
The two results are not inconsistent because the forces
that produce them act in orthogonal directions.
The tangential components of the stress at the liquid/vapor interface are what contribute to $F_s$ while changes in the component normal to the interface determine $P_{\rho \rho}$.
In macroscopic models the change from zero pressure outside the
drop to $P_{\rho\rho}$ within the drop is mediated by the tension in the interface.
This is the only possibility when all length scales are larger than the
interface width.
For $h<4\sigma$ there are direct interactions between solid atoms outside
the drop and fluid atoms near the center of the meniscus
that can replace the tension from the interface.
Indeed these interactions lead to the tension at the contact line in a larger
drop and the change in $P_{\rho \rho}$ there.
This explains why the radial stress can continue to follow
the macroscopic prediction even in extremely thin films.

\section{Conclusions}

The simulations presented here show that the macroscopic theory of 
capillarity works down to surprisingly small scales.
Within statistical uncertainties ($\sim 0.1\sigma$ and 1\%),
the meniscus shape and forces agree with the continuum prediction down
to $h\sim 8\sigma$ ($\sim 4~{\rm nm}$).
Below that scale, molecular layering produces strong oscillations in the force
that are not captured by macroscopic theory.
The shape of the liquid bridge and the Young-Laplace equation remain
accurate to even smaller scales ($h \sim 3$-$4\sigma$).

The shape of the interface remains consistent with macroscopic theory as
long as the radius of curvature is much larger than the molecular diameter.
Fig. \ref{InterfaceShape} shows good agreement even for $h=4.1\sigma$ and $\theta=12^\circ$ where $|r_1| = h/(2 \cos \theta) \sim 2.1\sigma$.
As $h$ decreases further, layering introduces discreteness in the
interface profile and limits the decrease in radius.
In the limit of one or two layers, the radius of the interface must
always be effectively independent of height, implying zero vertical
curvature and 
$\theta \sim 90^\circ$.
Since the surface tension contribution to the capillary force
$F_s \propto \sin{\theta}$,
there is a large increase in the magnitude of $F_s$
for fluids with equilibrium angles near $0^\circ$ (Fig. \ref{ForTotPlot}(f))
or $180^\circ$.

Detailed analysis of the local pressure tensor allowed us to isolate the trends in different contributions to the capillary force.
When $h \lesssim 8 \sigma$ the pressure inside the liquid bridge becomes anisotropic.
The in-plane component $P_{\rho\rho}$ is consistent with the macroscopic
solution for the Laplace pressure $\Delta p$ for $h$ down to $3$-$4\sigma$,
even when layering prevents the meniscus from following the high predicted
curvature.
For these very thin films, direct interactions with the wall span the entire
interface and replace the interfacial tension assumed in the Young-Laplace equation.
In contrast, the normal component $P_{zz}$ does not follow the macroscopic
solution. It has an additional contribution that
oscillates rapidly with $h$ and is large enough to change the sign
of the pressure as well as the magnitude.

The difference between normal and in-plane pressures, $\Delta p_d = P_{zz}-\Delta p$, represents a disjoining pressure
from changes in free energy per unit area with surface separation.
The oscillation period corresponds to the spacing between molecular layers that form near
the walls and span the system for $h \lesssim 8\sigma$.
For the short range interactions used here, $\Delta p_d$ is nearly independent
of changes in wall interactions that change $\theta$ from $75^\circ$ to $12^\circ$ (Fig. \ref{DisjoiningPressure}).
This indicates that $\Delta p_d$ is predominantly from entropy associated with layer packing at a given wall spacing.
The disjoining pressure is nearly independent of drop volume as long as the contact radius $a >> h$.

It is useful to compare the results from the simulations presented here 
to those in our previous study of capillary adhesion in the sphere-on-flat geometry.\cite{cheng14pre} 
This earlier paper also found that macroscopic theory described the interface
shape but did not study menisci with heights less than $10\sigma$ because the curvature of the sphere led to large separations at the liquid/vapor interface. 
The most significant difference between the two studies is
that the disjoining pressure for curved surfaces
always had an additional repulsive component that increased
with decreasing $h$ and sphere radius.
This repulsion is not present for the flat surfaces used here.
The increase in the magnitude of $\Delta p_d$
with increasing curvature confirms that it represents an additional entropic
cost associated with changing the number of layers as curvature changes the local
separation between the surfaces.

Experiments show that the range of the disjoining pressure can vary greatly
with molecular structure and interactions.\cite{horn81,toxvaerd81,magda85,degennes85,thompson90a,white03,bowles06,dutka14}
Our results suggest that the capillary force will be consistent with macroscopic
theory until the disjoining pressure is significant.
If the disjoining pressure is added to the macroscopic theory, the result
should remain accurate until $h$ approaches the molecular diameter or
the thickness of the liquid/vapor interface.
The two are the same in our system but the interface thickness may be larger,
for example near a critical point.
It will be interesting to test these ideas and to include the effect
of surface roughness which may increase repulsion and lead to variations
in the local contact angle.
Studies of plates with asymmetric wetting properties would
also allow tests of macroscopic predictions,
such as work that finds attraction whenever
the sum of the contact angles on the solid plates is less than 180 degrees.\cite{deSouza08a,deSouza08b}

\section*{Acknowledgement}
This material is based upon work supported by the National Science Foundation
under Grant No. DMR-1411144.


\begin{mcitethebibliography}{55}
\providecommand*\natexlab[1]{#1}
\providecommand*\mciteSetBstSublistMode[1]{}
\providecommand*\mciteSetBstMaxWidthForm[2]{}
\providecommand*\mciteBstWouldAddEndPuncttrue
  {\def\EndOfBibitem{\unskip.}}
\providecommand*\mciteBstWouldAddEndPunctfalse
  {\let\EndOfBibitem\relax}
\providecommand*\mciteSetBstMidEndSepPunct[3]{}
\providecommand*\mciteSetBstSublistLabelBeginEnd[3]{}
\providecommand*\EndOfBibitem{}
\mciteSetBstSublistMode{f}
\mciteSetBstMaxWidthForm{subitem}{(\alph{mcitesubitemcount})}
\mciteSetBstSublistLabelBeginEnd
  {\mcitemaxwidthsubitemform\space}
  {\relax}
  {\relax}

\bibitem[Hornbaker \latin{et~al.}(1997)Hornbaker, Albert, Albert, Barab\'{a}si,
  and Schiffer]{hornbaker97}
Hornbaker,~D.~J.; Albert,~R.; Albert,~I.; Barab\'{a}si,~A.~L.; Schiffer,~P.
  What Keeps Sandcastles Standing? \emph{Nature} \textbf{1997}, \emph{387},
  765--765\relax
\mciteBstWouldAddEndPuncttrue
\mciteSetBstMidEndSepPunct{\mcitedefaultmidpunct}
{\mcitedefaultendpunct}{\mcitedefaultseppunct}\relax
\EndOfBibitem
\bibitem[Weiss(2000)]{weiss00}
Weiss,~P. The Little Engines That Couldn't: Tired of Grinding Their Gears,
  Micromachine Researchers Turn to Surface Science. \emph{Science News}
  \textbf{2000}, \emph{158}, 56--58\relax
\mciteBstWouldAddEndPuncttrue
\mciteSetBstMidEndSepPunct{\mcitedefaultmidpunct}
{\mcitedefaultendpunct}{\mcitedefaultseppunct}\relax
\EndOfBibitem
\bibitem[Rabinovich \latin{et~al.}(2002)Rabinovich, Esayanur, Johanson, Adler,
  and Moudgil]{rabinovich02}
Rabinovich,~Y.~I.; Esayanur,~M.~S.; Johanson,~K.~D.; Adler,~J.~J.;
  Moudgil,~B.~M. Measurement of Oil-mediated Particle Adhesion to a Silica
  Substrate by Atomic Force Microscopy. \emph{J. Adhesion Sci. Technol.}
  \textbf{2002}, \emph{16}, 887--903\relax
\mciteBstWouldAddEndPuncttrue
\mciteSetBstMidEndSepPunct{\mcitedefaultmidpunct}
{\mcitedefaultendpunct}{\mcitedefaultseppunct}\relax
\EndOfBibitem
\bibitem[Herminghaus(2005)]{herminghaus05}
Herminghaus,~S. Dynamics of Wet Granular Matter. \emph{Adv. Phys.}
  \textbf{2005}, \emph{54}, 221--261\relax
\mciteBstWouldAddEndPuncttrue
\mciteSetBstMidEndSepPunct{\mcitedefaultmidpunct}
{\mcitedefaultendpunct}{\mcitedefaultseppunct}\relax
\EndOfBibitem
\bibitem[Scheel \latin{et~al.}(2008)Scheel, Seemann, Brinkmann, Michiel,
  Sheppard, Breidenbach, and Herminghaus]{scheel08}
Scheel,~M.; Seemann,~R.; Brinkmann,~M.; Michiel,~M.~D.; Sheppard,~A.;
  Breidenbach,~B.; Herminghaus,~S. Morphological Clues to Wet Granular Pile
  Stability. \emph{Nature Mater.} \textbf{2008}, \emph{7}, 189--193\relax
\mciteBstWouldAddEndPuncttrue
\mciteSetBstMidEndSepPunct{\mcitedefaultmidpunct}
{\mcitedefaultendpunct}{\mcitedefaultseppunct}\relax
\EndOfBibitem
\bibitem[Riedo \latin{et~al.}(2002)Riedo, L\'evy, and Brune]{riedo02}
Riedo,~E.; L\'evy,~F.; Brune,~H. Kinetics of Capillary Condensation in
  Nanoscopic Sliding Friction. \emph{Phys. Rev. Lett.} \textbf{2002},
  \emph{88}, 185505\relax
\mciteBstWouldAddEndPuncttrue
\mciteSetBstMidEndSepPunct{\mcitedefaultmidpunct}
{\mcitedefaultendpunct}{\mcitedefaultseppunct}\relax
\EndOfBibitem
\bibitem[Charlaix and Ciccotti(2010)Charlaix, and Ciccotti]{charlaix10}
Charlaix,~E.; Ciccotti,~M. In \emph{Handbook of Nanophysics: Principles and
  Methods}; Sattler,~K., Ed.; CRC Press: Boca Raton, FL, 2010\relax
\mciteBstWouldAddEndPuncttrue
\mciteSetBstMidEndSepPunct{\mcitedefaultmidpunct}
{\mcitedefaultendpunct}{\mcitedefaultseppunct}\relax
\EndOfBibitem
\bibitem[Maboudian \latin{et~al.}(2002)Maboudian, Ashurst, and
  Carraro]{maboudian02}
Maboudian,~R.; Ashurst,~W.~R.; Carraro,~C. Tribological Challenges in
  Micromechanical Systems. \emph{Trib. Lett.} \textbf{2002}, \emph{12},
  95--100\relax
\mciteBstWouldAddEndPuncttrue
\mciteSetBstMidEndSepPunct{\mcitedefaultmidpunct}
{\mcitedefaultendpunct}{\mcitedefaultseppunct}\relax
\EndOfBibitem
\bibitem[Rowlinson and Widom(1989)Rowlinson, and Widom]{rowlinson89}
Rowlinson,~J.~S.; Widom,~B. \emph{Molecular Theory of Capillarity}; Oxford:
  Oxford, 1989\relax
\mciteBstWouldAddEndPuncttrue
\mciteSetBstMidEndSepPunct{\mcitedefaultmidpunct}
{\mcitedefaultendpunct}{\mcitedefaultseppunct}\relax
\EndOfBibitem
\bibitem[Lambert \latin{et~al.}(2008)Lambert, Chau, and Delchambre]{lambert08}
Lambert,~P.; Chau,~A.; Delchambre,~A. Comparison between Two Capillary Forces
  Models. \emph{Langmuir} \textbf{2008}, \emph{24}, 3157--3163\relax
\mciteBstWouldAddEndPuncttrue
\mciteSetBstMidEndSepPunct{\mcitedefaultmidpunct}
{\mcitedefaultendpunct}{\mcitedefaultseppunct}\relax
\EndOfBibitem
\bibitem[Fisher and Israelachvili(1979)Fisher, and Israelachvili]{fisher79}
Fisher,~L.~R.; Israelachvili,~J.~N. Direct Experimental Verification of the
  Kelvin Equation for Capillary Condensation. \emph{Naturale} \textbf{1979},
  \emph{277}, 548--549\relax
\mciteBstWouldAddEndPuncttrue
\mciteSetBstMidEndSepPunct{\mcitedefaultmidpunct}
{\mcitedefaultendpunct}{\mcitedefaultseppunct}\relax
\EndOfBibitem
\bibitem[Fisher and Israelachvili(1981)Fisher, and Israelachvili]{fisher81a}
Fisher,~L.~R.; Israelachvili,~J.~N. Experimental Studies on the Applicability
  of the Kelvin Equation to Highly Curved Concave Menisci. \emph{J. Colloid
  Interface Sci.} \textbf{1981}, \emph{80}, 528--541\relax
\mciteBstWouldAddEndPuncttrue
\mciteSetBstMidEndSepPunct{\mcitedefaultmidpunct}
{\mcitedefaultendpunct}{\mcitedefaultseppunct}\relax
\EndOfBibitem
\bibitem[Fisher \latin{et~al.}(1981)Fisher, Gamble, and Middlehurst]{fisher81b}
Fisher,~L.~R.; Gamble,~R.~A.; Middlehurst,~J. The Kelvin Equation and the
  Capillary Condensation of Water. \emph{Nature} \textbf{1981}, \emph{290},
  575--576\relax
\mciteBstWouldAddEndPuncttrue
\mciteSetBstMidEndSepPunct{\mcitedefaultmidpunct}
{\mcitedefaultendpunct}{\mcitedefaultseppunct}\relax
\EndOfBibitem
\bibitem[Kohonen and Christenson(2000)Kohonen, and Christenson]{kohonen00}
Kohonen,~M.~M.; Christenson,~H.~K. Capillary Condensation of Water between
  Rinsed Mica Surfaces. \emph{Langmuir} \textbf{2000}, \emph{16},
  7285--7288\relax
\mciteBstWouldAddEndPuncttrue
\mciteSetBstMidEndSepPunct{\mcitedefaultmidpunct}
{\mcitedefaultendpunct}{\mcitedefaultseppunct}\relax
\EndOfBibitem
\bibitem[Thompson \latin{et~al.}(1993)Thompson, Brinckerhoff, and
  Robbins]{thompson93b}
Thompson,~P.~A.; Brinckerhoff,~W.~B.; Robbins,~M.~O. Microscopic Studies of
  Static and Dynamic Contact Angles. \emph{J. Adhesion Sci. Technol.}
  \textbf{1993}, \emph{7}, 535--554\relax
\mciteBstWouldAddEndPuncttrue
\mciteSetBstMidEndSepPunct{\mcitedefaultmidpunct}
{\mcitedefaultendpunct}{\mcitedefaultseppunct}\relax
\EndOfBibitem
\bibitem[Bresme and Quirke(1998)Bresme, and Quirke]{bresme98}
Bresme,~F.; Quirke,~N. Computer Simuation Study of the Wetting Behavior and
  Line Tensions of Nanometer Size Particulates at a Liquid-vapor Interface.
  \emph{Phys. Rev. Lett.} \textbf{1998}, \emph{80}, 3791--3794\relax
\mciteBstWouldAddEndPuncttrue
\mciteSetBstMidEndSepPunct{\mcitedefaultmidpunct}
{\mcitedefaultendpunct}{\mcitedefaultseppunct}\relax
\EndOfBibitem
\bibitem[Takahashi and Morita(2013)Takahashi, and Morita]{takahashi13}
Takahashi,~H.; Morita,~A. A Molecular Dynamics Study on Inner Pressure of
  Microbubbles in Liquid Argon and Water. \emph{Chem. Phys. Lett.}
  \textbf{2013}, \emph{573}, 35 -- 40\relax
\mciteBstWouldAddEndPuncttrue
\mciteSetBstMidEndSepPunct{\mcitedefaultmidpunct}
{\mcitedefaultendpunct}{\mcitedefaultseppunct}\relax
\EndOfBibitem
\bibitem[Cheng and Robbins(2014)Cheng, and Robbins]{cheng14pre}
Cheng,~S.; Robbins,~M.~O. Capillary Adhesion at the Nanometer Scale.
  \emph{Phys. Rev. E} \textbf{2014}, \emph{89}, 062402\relax
\mciteBstWouldAddEndPuncttrue
\mciteSetBstMidEndSepPunct{\mcitedefaultmidpunct}
{\mcitedefaultendpunct}{\mcitedefaultseppunct}\relax
\EndOfBibitem
\bibitem[Mastrangeli(2015)]{mastrangeli15}
Mastrangeli,~M. The Fluid Joint: The Soft Spot of Micro- and Nanosystems.
  \emph{Adv. Mater.} \textbf{2015}, \emph{27}, 4254--4272\relax
\mciteBstWouldAddEndPuncttrue
\mciteSetBstMidEndSepPunct{\mcitedefaultmidpunct}
{\mcitedefaultendpunct}{\mcitedefaultseppunct}\relax
\EndOfBibitem
\bibitem[Kremer and Grest(1990)Kremer, and Grest]{kremer90}
Kremer,~K.; Grest,~G.~S. Dynamics of Entangled Linear Polymer Melts: A
  Molecular-Dynamics Simulation. \emph{J. Chem. Phys.} \textbf{1990},
  \emph{92}, 5057--5086\relax
\mciteBstWouldAddEndPuncttrue
\mciteSetBstMidEndSepPunct{\mcitedefaultmidpunct}
{\mcitedefaultendpunct}{\mcitedefaultseppunct}\relax
\EndOfBibitem
\bibitem[Kr\"oger \latin{et~al.}(1993)Kr\"oger, Loose, and Hess]{kroger93}
Kr\"oger,~M.; Loose,~W.; Hess,~S. Rheology and Structural Changes of Polymer
  Melts via Nonequilibrium Molecular Dynamics. \emph{J. Rheology}
  \textbf{1993}, \emph{37}, 1057--1079\relax
\mciteBstWouldAddEndPuncttrue
\mciteSetBstMidEndSepPunct{\mcitedefaultmidpunct}
{\mcitedefaultendpunct}{\mcitedefaultseppunct}\relax
\EndOfBibitem
\bibitem[Hur \latin{et~al.}(2000)Hur, Shaqfeh, and Larson]{larson00}
Hur,~J.~S.; Shaqfeh,~E. S.~G.; Larson,~R.~G. Brownian Dynamics Simulations of
  Single DNA Molecules in Shear Flow. \emph{J. Rheology} \textbf{2000},
  \emph{44}, 713--742\relax
\mciteBstWouldAddEndPuncttrue
\mciteSetBstMidEndSepPunct{\mcitedefaultmidpunct}
{\mcitedefaultendpunct}{\mcitedefaultseppunct}\relax
\EndOfBibitem
\bibitem[Rottler and Robbins(2003)Rottler, and Robbins]{rottler03c}
Rottler,~J.; Robbins,~M.~O. Shear Yielding of Amorphous Glassy Solids: Effect
  of Temperature and Strain Rate. \emph{Phys. Rev. E} \textbf{2003}, \emph{68},
  011507\relax
\mciteBstWouldAddEndPuncttrue
\mciteSetBstMidEndSepPunct{\mcitedefaultmidpunct}
{\mcitedefaultendpunct}{\mcitedefaultseppunct}\relax
\EndOfBibitem
\bibitem[Kr\"{o}ger(2004)]{kroger04}
Kr\"{o}ger,~M. Simple Models for Complex Nonequilibrium Fluids. \emph{Phys.
  Rep.} \textbf{2004}, \emph{390}, 453 -- 551\relax
\mciteBstWouldAddEndPuncttrue
\mciteSetBstMidEndSepPunct{\mcitedefaultmidpunct}
{\mcitedefaultendpunct}{\mcitedefaultseppunct}\relax
\EndOfBibitem
\bibitem[Cheng \latin{et~al.}(2010)Cheng, Luan, and Robbins]{cheng10pre}
Cheng,~S.; Luan,~B.~Q.; Robbins,~M.~O. Contact and Friction of Nanoasperities:
  Effects of Adsorbed Monolayers. \emph{Phys. Rev. E} \textbf{2010}, \emph{81},
  016102\relax
\mciteBstWouldAddEndPuncttrue
\mciteSetBstMidEndSepPunct{\mcitedefaultmidpunct}
{\mcitedefaultendpunct}{\mcitedefaultseppunct}\relax
\EndOfBibitem
\bibitem[LAM()]{LAMMPS}
{http://lammps.sandia.gov/}\relax
\mciteBstWouldAddEndPuncttrue
\mciteSetBstMidEndSepPunct{\mcitedefaultmidpunct}
{\mcitedefaultendpunct}{\mcitedefaultseppunct}\relax
\EndOfBibitem
\bibitem[Buchholz \latin{et~al.}(2002)Buchholz, Paul, Varnik, and
  Binder]{buchholz02}
Buchholz,~J.; Paul,~W.; Varnik,~F.; Binder,~K. Cooling Rate Dependence of the
  Glass Transition Temperature of Polymer Melts: Molecular Dynamics Study.
  \emph{J. Chem. Phys.} \textbf{2002}, \emph{117}, 7364--7372\relax
\mciteBstWouldAddEndPuncttrue
\mciteSetBstMidEndSepPunct{\mcitedefaultmidpunct}
{\mcitedefaultendpunct}{\mcitedefaultseppunct}\relax
\EndOfBibitem
\bibitem[Style and Dufresne(2012)Style, and Dufresne]{style12}
Style,~R.~W.; Dufresne,~E.~R. Static Wetting on Deformable Substrates{,} from
  Liquids to Soft Solids. \emph{Soft Matter} \textbf{2012}, \emph{8},
  7177--7184\relax
\mciteBstWouldAddEndPuncttrue
\mciteSetBstMidEndSepPunct{\mcitedefaultmidpunct}
{\mcitedefaultendpunct}{\mcitedefaultseppunct}\relax
\EndOfBibitem
\bibitem[Style \latin{et~al.}(2013)Style, Boltyanskiy, Che, Wettlaufer, Wilen,
  and Dufresne]{style13}
Style,~R.~W.; Boltyanskiy,~R.; Che,~Y.; Wettlaufer,~J.~S.; Wilen,~L.~A.;
  Dufresne,~E.~R. Universal Deformation of Soft Substrates Near a Contact Line
  and the Direct Measurement of Solid Surface Stresses. \emph{Phys. Rev. Lett.}
  \textbf{2013}, \emph{110}, 066103\relax
\mciteBstWouldAddEndPuncttrue
\mciteSetBstMidEndSepPunct{\mcitedefaultmidpunct}
{\mcitedefaultendpunct}{\mcitedefaultseppunct}\relax
\EndOfBibitem
\bibitem[Amirfazli and Neumann(2004)Amirfazli, and Neumann]{amirfazli04}
Amirfazli,~A.; Neumann,~A. Status of the Three-phase Line Tension: A Review.
  \emph{Adv. Colloid Interface Sci.} \textbf{2004}, \emph{110}, 121 --
  141\relax
\mciteBstWouldAddEndPuncttrue
\mciteSetBstMidEndSepPunct{\mcitedefaultmidpunct}
{\mcitedefaultendpunct}{\mcitedefaultseppunct}\relax
\EndOfBibitem
\bibitem[Cheng \latin{et~al.}(2011)Cheng, Lechman, Plimpton, and
  Grest]{cheng11jcp}
Cheng,~S.; Lechman,~J.~B.; Plimpton,~S.~J.; Grest,~G.~S. Evaporation of
  Lennard-Jones Fluids. \emph{J. Chem. Phys.} \textbf{2011}, \emph{134},
  224704\relax
\mciteBstWouldAddEndPuncttrue
\mciteSetBstMidEndSepPunct{\mcitedefaultmidpunct}
{\mcitedefaultendpunct}{\mcitedefaultseppunct}\relax
\EndOfBibitem
\bibitem[Orr \latin{et~al.}(1975)Orr, Scriven, and Rivas]{orr75}
Orr,~F.~M.; Scriven,~L.~E.; Rivas,~A.~P. Pendular Rings between Solids:
  Meniscus Properties and Capillary Force. \emph{J. Fluid Mech.} \textbf{1975},
  \emph{67}, 723--742\relax
\mciteBstWouldAddEndPuncttrue
\mciteSetBstMidEndSepPunct{\mcitedefaultmidpunct}
{\mcitedefaultendpunct}{\mcitedefaultseppunct}\relax
\EndOfBibitem
\bibitem[Nordholm and Haymet(1980)Nordholm, and Haymet]{nordholm80}
Nordholm,~S.; Haymet,~A. D.~J. Generalized van der Waals Theory. I Basic
  Formulation and Application to Uniform Fluids. \emph{Aust. J. Chem.}
  \textbf{1980}, \emph{33}, 2013--2027\relax
\mciteBstWouldAddEndPuncttrue
\mciteSetBstMidEndSepPunct{\mcitedefaultmidpunct}
{\mcitedefaultendpunct}{\mcitedefaultseppunct}\relax
\EndOfBibitem
\bibitem[Toxvaerd(1981)]{toxvaerd81}
Toxvaerd,~S. The Structure and Thermodynamics of a Solid-fluid Interface.
  \emph{J. Chem. Phys.} \textbf{1981}, \emph{74}, 1998--2008\relax
\mciteBstWouldAddEndPuncttrue
\mciteSetBstMidEndSepPunct{\mcitedefaultmidpunct}
{\mcitedefaultendpunct}{\mcitedefaultseppunct}\relax
\EndOfBibitem
\bibitem[Magda \latin{et~al.}(1985)Magda, Tirrell, and Davis]{magda85}
Magda,~J.; Tirrell,~M.; Davis,~H.~T. Molecular Dynamics of Narrow,
  Liquid-filled Pores. \emph{J. Chem. Phys.} \textbf{1985}, \emph{83},
  1888--1901\relax
\mciteBstWouldAddEndPuncttrue
\mciteSetBstMidEndSepPunct{\mcitedefaultmidpunct}
{\mcitedefaultendpunct}{\mcitedefaultseppunct}\relax
\EndOfBibitem
\bibitem[Bitsanis and Hadziioannou(1990)Bitsanis, and
  Hadziioannou]{bitsanis90a}
Bitsanis,~I.; Hadziioannou,~G. Molecular Dynamics Simulations of the Structure
  and Dynamics of Confined Polymer Melts. \emph{J. Chem. Phys.} \textbf{1990},
  \emph{92}, 3827--3847\relax
\mciteBstWouldAddEndPuncttrue
\mciteSetBstMidEndSepPunct{\mcitedefaultmidpunct}
{\mcitedefaultendpunct}{\mcitedefaultseppunct}\relax
\EndOfBibitem
\bibitem[Gao \latin{et~al.}(1997)Gao, Luedtke, and Landman]{gao97c}
Gao,~J.; Luedtke,~W.~D.; Landman,~U. Layering Transitions and Dynamics of
  Confined Liquid Fims. \emph{Phys. Rev. Lett.} \textbf{1997}, \emph{79},
  705--708\relax
\mciteBstWouldAddEndPuncttrue
\mciteSetBstMidEndSepPunct{\mcitedefaultmidpunct}
{\mcitedefaultendpunct}{\mcitedefaultseppunct}\relax
\EndOfBibitem
\bibitem[Thompson and Robbins(1990)Thompson, and Robbins]{thompson90a}
Thompson,~P.~A.; Robbins,~M.~O. Shear Flow Near Solids: Epitaxial Order and
  Flow Boundary Conditions. \emph{Phys. Rev. A} \textbf{1990}, \emph{41},
  6830--6837\relax
\mciteBstWouldAddEndPuncttrue
\mciteSetBstMidEndSepPunct{\mcitedefaultmidpunct}
{\mcitedefaultendpunct}{\mcitedefaultseppunct}\relax
\EndOfBibitem
\bibitem[Jasnow and Vi\~{n}als(1996)Jasnow, and Vi\~{n}als]{jasnow96}
Jasnow,~D.; Vi\~{n}als,~J. Coarse-grained Description of Thermo-capillary Flow.
  \emph{Phys. Fluids} \textbf{1996}, \emph{8}, 660--669\relax
\mciteBstWouldAddEndPuncttrue
\mciteSetBstMidEndSepPunct{\mcitedefaultmidpunct}
{\mcitedefaultendpunct}{\mcitedefaultseppunct}\relax
\EndOfBibitem
\bibitem[Denniston and Robbins(2004)Denniston, and Robbins]{denniston04}
Denniston,~C.; Robbins,~M.~O. Mapping Molecular Models to Continuum Theories
  for Partially Miscible Fluids. \emph{Phys. Rev. E} \textbf{2004}, \emph{69},
  021505\relax
\mciteBstWouldAddEndPuncttrue
\mciteSetBstMidEndSepPunct{\mcitedefaultmidpunct}
{\mcitedefaultendpunct}{\mcitedefaultseppunct}\relax
\EndOfBibitem
\bibitem[Lacasse \latin{et~al.}(1998)Lacasse, Grest, and Levine]{lacasse98}
Lacasse,~M.-D.; Grest,~G.~S.; Levine,~A.~J. Capillary-wave and Chain-length
  Effects at Polymer/Polymer Interfaces. \emph{Phys. Rev. Lett.} \textbf{1998},
  \emph{80}, 309--312\relax
\mciteBstWouldAddEndPuncttrue
\mciteSetBstMidEndSepPunct{\mcitedefaultmidpunct}
{\mcitedefaultendpunct}{\mcitedefaultseppunct}\relax
\EndOfBibitem
\bibitem[Denniston and Robbins(2006)Denniston, and Robbins]{denniston06}
Denniston,~C.; Robbins,~M.~O. Matching Continuum Boundary Conditions to
  Molecular Dynamics Simulations for a Miscible Binary Fluid. \emph{J. Chem.
  Phys.} \textbf{2006}, \emph{125}, 214102\relax
\mciteBstWouldAddEndPuncttrue
\mciteSetBstMidEndSepPunct{\mcitedefaultmidpunct}
{\mcitedefaultendpunct}{\mcitedefaultseppunct}\relax
\EndOfBibitem
\bibitem[Horn and Israelachvili(1981)Horn, and Israelachvili]{horn81}
Horn,~R.~G.; Israelachvili,~J.~N. Direct Measurement of Structural Forces
  between Two Surfaces in a Nonpolar Liquid. \emph{J. Chem. Phys.}
  \textbf{1981}, \emph{75}, 1400--1412\relax
\mciteBstWouldAddEndPuncttrue
\mciteSetBstMidEndSepPunct{\mcitedefaultmidpunct}
{\mcitedefaultendpunct}{\mcitedefaultseppunct}\relax
\EndOfBibitem
\bibitem[Israelachvili \latin{et~al.}(1984)Israelachvili, Tirrell, Klein, and
  Almog]{israelachvili84}
Israelachvili,~J.~N.; Tirrell,~M.; Klein,~J.; Almog,~Y. Forces between Two
  Layers of Adsorbed Polystyrene Immersed in Cyclohexane below and above the
  $\theta$ Temperature. \emph{Macromolecules} \textbf{1984}, \emph{17},
  204--209\relax
\mciteBstWouldAddEndPuncttrue
\mciteSetBstMidEndSepPunct{\mcitedefaultmidpunct}
{\mcitedefaultendpunct}{\mcitedefaultseppunct}\relax
\EndOfBibitem
\bibitem[Jarvis \latin{et~al.}(2000)Jarvis, Uchihashi, Ishida, Tokumoto, and
  Nakayama]{jarvis00}
Jarvis,~S.~P.; Uchihashi,~T.; Ishida,~T.; Tokumoto,~H.; Nakayama,~Y. Local
  Solvation Shell Measurement in Water Using a Carbon Nanotube Probe. \emph{J.
  Phys. Chem. B} \textbf{2000}, \emph{104}, 6091--6094\relax
\mciteBstWouldAddEndPuncttrue
\mciteSetBstMidEndSepPunct{\mcitedefaultmidpunct}
{\mcitedefaultendpunct}{\mcitedefaultseppunct}\relax
\EndOfBibitem
\bibitem[Ko\ifmmode~\check{c}\else \v{c}\fi{}evar
  \latin{et~al.}(2001)Ko\ifmmode~\check{c}\else \v{c}\fi{}evar,
  Bor\ifmmode~\check{s}\else \v{s}\fi{}tnik, Mu\ifmmode \check{s}\else
  \v{s}\fi{}evi\ifmmode~\check{c}\else \v{c}\fi{}, and \ifmmode~\check{Z}\else
  \v{Z}\fi{}umer]{kocevar01}
Ko\ifmmode~\check{c}\else \v{c}\fi{}evar,~K.; Bor\ifmmode~\check{s}\else
  \v{s}\fi{}tnik,~A.; Mu\ifmmode \check{s}\else
  \v{s}\fi{}evi\ifmmode~\check{c}\else \v{c}\fi{},~I.; \ifmmode~\check{Z}\else
  \v{Z}\fi{}umer,~S. Capillary Condensation of a Nematic Liquid Crystal
  Observed by Force Spectroscopy. \emph{Phys. Rev. Lett.} \textbf{2001},
  \emph{86}, 5914--5917\relax
\mciteBstWouldAddEndPuncttrue
\mciteSetBstMidEndSepPunct{\mcitedefaultmidpunct}
{\mcitedefaultendpunct}{\mcitedefaultseppunct}\relax
\EndOfBibitem
\bibitem[Todd \latin{et~al.}(1995)Todd, Evans, and Daivis]{todd95}
Todd,~B.~D.; Evans,~D.~J.; Daivis,~P.~J. Pressure Tensor for Inhomogeneous
  Fluids. \emph{Phys. Rev. E} \textbf{1995}, \emph{52}, 1627--1638\relax
\mciteBstWouldAddEndPuncttrue
\mciteSetBstMidEndSepPunct{\mcitedefaultmidpunct}
{\mcitedefaultendpunct}{\mcitedefaultseppunct}\relax
\EndOfBibitem
\bibitem[Denniston and Robbins(2001)Denniston, and Robbins]{denniston01}
Denniston,~C.; Robbins,~M.~O. Molecular and Continuum Boundary Conditions for a
  Miscible Binary Fluid. \emph{Phys. Rev. Lett.} \textbf{2001}, \emph{87},
  178302\relax
\mciteBstWouldAddEndPuncttrue
\mciteSetBstMidEndSepPunct{\mcitedefaultmidpunct}
{\mcitedefaultendpunct}{\mcitedefaultseppunct}\relax
\EndOfBibitem
\bibitem[de~Gennes(1985)]{degennes85}
de~Gennes,~P.~G. Wetting: Statics and Dynamics. \emph{Rev. Mod. Phys.}
  \textbf{1985}, \emph{57}, 827--863\relax
\mciteBstWouldAddEndPuncttrue
\mciteSetBstMidEndSepPunct{\mcitedefaultmidpunct}
{\mcitedefaultendpunct}{\mcitedefaultseppunct}\relax
\EndOfBibitem
\bibitem[White(2003)]{white03}
White,~L.~R. The Contact Angle on an Elastic Substrate. 1. The Role of
  Disjoining Pressure in the Surface Mechanics. \emph{J. Colloid Interface
  Sci.} \textbf{2003}, \emph{258}, 82--96\relax
\mciteBstWouldAddEndPuncttrue
\mciteSetBstMidEndSepPunct{\mcitedefaultmidpunct}
{\mcitedefaultendpunct}{\mcitedefaultseppunct}\relax
\EndOfBibitem
\bibitem[Bowles \latin{et~al.}(2006)Bowles, Hsia, Jones, Schneider, and
  White]{bowles06}
Bowles,~A.~P.; Hsia,~Y.-T.; Jones,~P.~M.; Schneider,~J.~W.; White,~L.~R.
  Quasi-equilibrium {AFM} Measurement of Disjoining Pressure in Lubricant
  Nano-films I: Fomblin Z03 on Silica. \emph{Langmuir} \textbf{2006},
  \emph{22}, 11436--11446\relax
\mciteBstWouldAddEndPuncttrue
\mciteSetBstMidEndSepPunct{\mcitedefaultmidpunct}
{\mcitedefaultendpunct}{\mcitedefaultseppunct}\relax
\EndOfBibitem
\bibitem[Dutka and Napi\'{o}rkowski(2014)Dutka, and Napi\'{o}rkowski]{dutka14}
Dutka,~F.; Napi\'{o}rkowski,~M. The Influence of van der Waals Forces on
  Droplet Morphological Transitions and Solvation Forces in Nanochannels.
  \emph{J. Phys.: Condens. Matter} \textbf{2014}, \emph{26}, 035101\relax
\mciteBstWouldAddEndPuncttrue
\mciteSetBstMidEndSepPunct{\mcitedefaultmidpunct}
{\mcitedefaultendpunct}{\mcitedefaultseppunct}\relax
\EndOfBibitem
\bibitem[Souza \latin{et~al.}(2008)Souza, Gao, McCarthy, Arzt, and
  Crosby]{deSouza08a}
Souza,~E. J.~D.; Gao,~L.; McCarthy,~T.~J.; Arzt,~E.; Crosby,~A.~J. Effect of
  Contact Angle Hysteresis on the Measurement of Capillary Forces.
  \emph{Langmuir} \textbf{2008}, \emph{24}, 1391--1396\relax
\mciteBstWouldAddEndPuncttrue
\mciteSetBstMidEndSepPunct{\mcitedefaultmidpunct}
{\mcitedefaultendpunct}{\mcitedefaultseppunct}\relax
\EndOfBibitem
\bibitem[Souza \latin{et~al.}(2008)Souza, Brinkmann, Mohrdieck, Crosby, and
  Arzt]{deSouza08b}
Souza,~E. J.~D.; Brinkmann,~M.; Mohrdieck,~C.; Crosby,~A.~J.; Arzt,~E.
  Capillary Forces between Chemically Different Substrates. \emph{Langmuir}
  \textbf{2008}, \emph{24}, 10161--10168\relax
\mciteBstWouldAddEndPuncttrue
\mciteSetBstMidEndSepPunct{\mcitedefaultmidpunct}
{\mcitedefaultendpunct}{\mcitedefaultseppunct}\relax
\EndOfBibitem
\end{mcitethebibliography}
\providecommand{\latin}[1]{#1}
\providecommand*\mcitethebibliography{\thebibliography}
\csname @ifundefined\endcsname{endmcitethebibliography}
  {\let\endmcitethebibliography\endthebibliography}{}

\end{document}